\documentclass[prb,preprint,showpacs,superscriptaddress]{revtex4}
\usepackage{graphicx}
\begin{document}
\title{Disorder effects on the superconducting properties of BaFe$_{1.8}$Co$_{0.2}$As$_2$ single
crystals}

\author{M~Eisterer}
\author{M~Zehetmayer}
\author{H W Weber}
\address{Atomic Institute, Vienna University of Technology, Austria}

\author{J~Jiang}
\author{J~D~Weiss}
\author{A~Yamamoto}
\author{E~E~Hellstrom}
\address{National High Magnetic Field Laboratory, Florida State University, Tallahassee,
FL 32310, USA }


\begin{abstract}
Single crystals of superconducting BaFe$_{1.8}$Co$_{0.2}$As$_2$
were exposed to neutron irradiation in a fission reactor. The
introduced defects decrease the superconducting transition
temperature (by about 0.3\,K) and the upper critical field
anisotropy (e.g. from 2.8 to 2.5 at 22\,K) and enhance the
critical current densities by a factor of up to about 3. These
changes are discussed in the context of similar experiments on
other superconducting materials.

\end{abstract}


\maketitle

\newpage

\section{Introduction}

Neutron irradiation is a powerful tool for theoretical studies as
well as the optimization of superconductors
\cite{Swe78,Sau98,Wer00,Zeh02b,Zeh04,Zeh05,Kru07,Put07,Eis09b},
since electron scattering centers and/or pinning sites are
introduced. The effects can be investigated on the same sample
allowing comparisons with theoretical predictions for changes in
inter- and intra-band scattering \cite{Swe78,Kru07,Put07} or with
pinning models for uncorrelated, randomly distributed, nano-sized
pinning centers \cite{Wer00,Zeh04,Zeh05, Kru07}. Scattering seems
to be particularly interesting in the FeAs based materials since
they are multi-band superconductors, possibly with (presently
still unexplored) extended s-wave pairing symmetry \cite{Maz08}.
The changes in scattering and pinning also provide guidelines for
material optimization.

\section{Experimental\label{secexp}}

The BaFe$_{1.8}$Co$_{0.2}$As$_2$ crystals (typically
$1.4\times0.7\times0.1$\,mm$^3$) were prepared by the self-flux
method \cite{Sef08} at the National High Magnetic Field Laboratory
and irradiated at the Atomic Institute to a fast neutron fluence
of $4\times 10^{21}$\,m$^{-2}$. Defects are introduced either by
direct collisions of high energy neutrons with lattice atoms or by
nuclear reactions. All nuclei can capture neutrons followed by a
prompt gamma emission of a few MeV. The recoil energy is
sufficient to displace the emitting nucleus. The same holds for
the $\beta$-decay of $^{76}$As. However, many of the resulting
Frenkel pairs recombine quickly and the stable defects add to
those produced by direct collisions, whose density is also
unknown. Larger defects can be created only by fast neutrons
($E>0.1$\,MeV).

One crystal was mounted onto a rotating sample holder and
investigated in a superconducting 17\,T solenoid. A current of
300\,$\mu$A ($\sim 3$\,kA\,m$^{-2}$) was applied in order to
measure resistivity at various angles and magnetic fields while
cooling at a rate of 10\,K\,h$^{-1}$. The upper critical field was
evaluated using three different criteria: 90\,\% (onset), 50\,\%
(midpoint), and 10\,\% (offset) of the normal state resistivity,
which was extrapolated linearly from its behaviour just above the
transition temperature. The transition width, $\Delta
T_\mathrm{c}$, refers to the temperature difference between the
onset and the offset. All results refer to the 50\,\% criterion in
the following, although the same analysis was also made on the
basis of the 90\,\% and 10\,\% criteria. No qualitative influence
on the results was found (cf. Fig.~\ref{FiggammaT}), but data
scattering was smallest for the 50\,\% criterion, since the
transition is steepest there.

A second crystal was characterized magnetically. Its transition
temperature was measured in a SQUID using ac-mode with a field
amplitude of 0.1\,mT. Irreversible properties were studied from
magnetization loops - $m(H_\mathrm{a})$ - with $H_\mathrm{a}
\parallel c$ measured in a VSM at fields up to 5\,T and a field
sweep rate of 10$^{-2}$ T\,s$^{-1}$. The critical current density,
$J_\mathrm{c}$, was evaluated by applying Bean's model for
rectangular samples (e.g. Ref.~\onlinecite{Zeh04}).

\section{Resistivity and transition temperature\label{secrestc}}

\begin{figure} \centering \includegraphics[clip,width=0.4\textwidth]{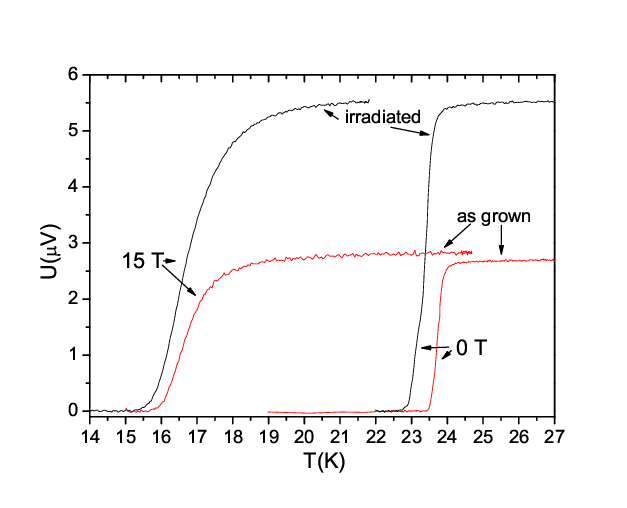}
\caption{Resistive transitions at 0 and 15 T ($B\parallel c$)
prior to and after neutron irradiation.} \label{FigTran}
\end{figure}

The transition temperature was found to be 23.75\,K with a small
transition width of 0.35\,K (Fig.~\ref{FigTran}), increasing to
only about 1.5\,K at 15\,T ($B\parallel c$, perpendicular to the
FeAs layers). $T_\mathrm{c}$ decreased to 23.4\,K after
irradiation and $\Delta T_\mathrm{c}$ at zero field increased to
0.6\,K. The decrease in $T_\mathrm{c}$ was confirmed by ac
susceptibility measurements on the second crystal, form $\simeq
23.4$\,K to 23.1\,K. A decrease in $T_\mathrm{c}$ was recently
predicted to result from impurity scattering in extended s-wave
superconductors \cite{Ban09}. However, the decrease in
$T_\mathrm{c}$ after neutron irradiation is a rather common
feature observed in many different materials, such as the cuprates
\cite{Sau98, Zeh05}, MgB$_2$ \cite{Kru07} and the A-15-compounds
\cite{Swe78}, since it can also be caused by d-wave
superconductivity \cite{Mil88}, interband scattering \cite{Put07},
a change in the electronic density of states \cite{Put07} or by a
reduction in anisotropy \cite{Mil88}.

An enhanced density of impurity scattering centres after neutron
irradiation is evidenced by the residual resistivity ratio,
$RRR=\rho_n(300\,\mathrm{K})/\rho_n(25\,\mathrm{K})$, which
decreases from 2.2 to 1.9. Since the resistivity is only weakly
temperature dependent above $T_\mathrm{c}$,
$\rho_n(25\,\mathrm{K})$ represents a reasonable estimate for the
residual resistivity $\rho_0$, which seems to increase by a factor
of about 2 in Fig.~\ref{FigTran}. We did not calculate the
absolute values of the resistivity due to large uncertainties in
the determination of the crystal geometry. Also the absolute
change in voltage is not reliable since we did not use the same
contacts before and after irradiation, therefore, the distance
between the voltage taps slightly changed and could not be
determined exactly because of the finite contact area. Only the
$RRR$ is independent of the actual geometry.

\section{Upper critical field and anisotropy\label{secbc2}}

\begin{figure} \centering \includegraphics[clip,width=0.4\textwidth]{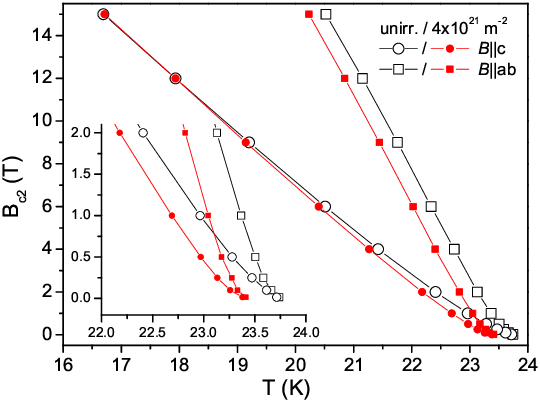}
\caption{The upper critical field, for the two main field
orientations, $B_\mathrm{c2}^\perp(T)$ and
$B_\mathrm{c2}^\parallel(T)$. The irradiation increases the slope
only for $B\parallel c$.} \label{FigBc2}
\end{figure}

The upper critical field for both main field orientations is
plotted as a function of temperature in Fig.~\ref{FigBc2}. While
$B_\mathrm{c2}^{\parallel}(T)$ (field parallel to the ab planes)
shifts to lower temperatures without a significant change in
slope, $B_\mathrm{c2}^{\perp}(T)$ becomes steeper after
irradiation leading to an enhancement at low temperatures. This is
expected from a reduction of the electron mean free path due to
impurity scattering in s-wave superconductors. In d-wave
superconductors a \emph{decrease} of $B_\mathrm{c2}$ is predicted
theoretically \cite{Won94}. We are not aware of the corresponding
prediction for extended s-wave pairing.

\begin{figure} \centering \includegraphics[clip,width=0.4\textwidth]{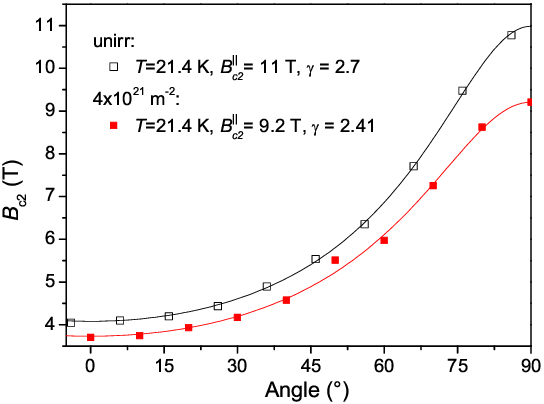}
\caption{The angular dependence of the upper critical field agrees
with the prediction of anisotropic Ginzburg Landau theory (solid
lines) prior to and after the irradiation} \label{Figang}
\end{figure}

The angular dependence of the upper critical field is plotted in
Fig.~\ref{Figang}. $B_\mathrm{c2}(\theta)$ at a fixed temperature
was obtained from linear interpolation between the values measured
at fixed fields (cf. Fig.~\ref{FigBc2}). This does not induce a
significant systematic error, since the curvature of
$B_\mathrm{c2}(T)$ is small except for a small temperature region
near $T_\mathrm{c}$, where the number of measured points was high
anyway.

\begin{figure} \centering \includegraphics[clip,width=0.4\textwidth]{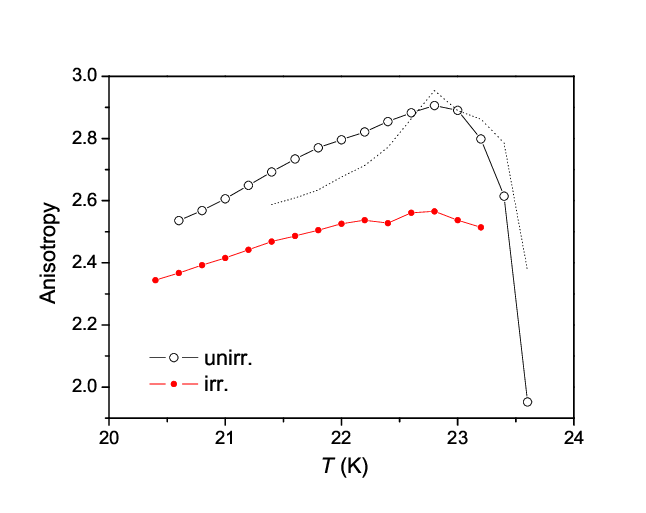}
\caption{The upper critical field anisotropy increases with
temperature, except in the close vicinity of $T_\mathrm{c}$. The
irradiation reduces the anisotropy. Dotted line graph represents
$\gamma(T)$ of the as grown sample evaluated by the 90\% criterion
for comparison. (The other data refer to the 50\% criterion.)}
\label{FiggammaT}
\end{figure}

The data are in excellent agreement with anisotropic Ginzburg
Landau theory,
$B_{c2}(\theta)={B_{c2}^{\parallel}}({{\gamma^{2}\cos^{2}(\theta)+\sin^{2}(\theta)}})^{-0.5}$.
The upper critical field anisotropy, $\gamma$, was determined by
fitting this relation to the experimental data. The anisotropy has
a maximum of 2.9 at 22.8\,K, which is reduced to 2.55 after
irradiation (Fig.~\ref{FiggammaT}). A reduction in $\gamma$ upon
neutron irradiation was also found in the cuprates \cite{Zeh05}
and in MgB$_2$ \cite{Kru07}. The pronounced drop of $\gamma$ near
$T_\mathrm{c}$ can be directly observed in the inset of
Fig.~\ref{FigBc2} and is obviously related to the curvature of
$B_\mathrm{c2}(T)$.

The anisotropy agrees with previous reports
\cite{Ni08,Tan09,Yam09}. The present experiment demonstrates that
$\gamma$ is sensitive to impurity scattering, which will result in
sample to sample variations. Note that electron scattering by the
doping atoms (Co in this case) cannot explain the doping
dependence of the anisotropy, since an increase of $\gamma$ was
found at higher concentrations of doping atoms \cite{Ni08}.

\section{Critical currents\label{secJc}}

\begin{figure}
    \centering
    \includegraphics[clip, width =0.35\textwidth]{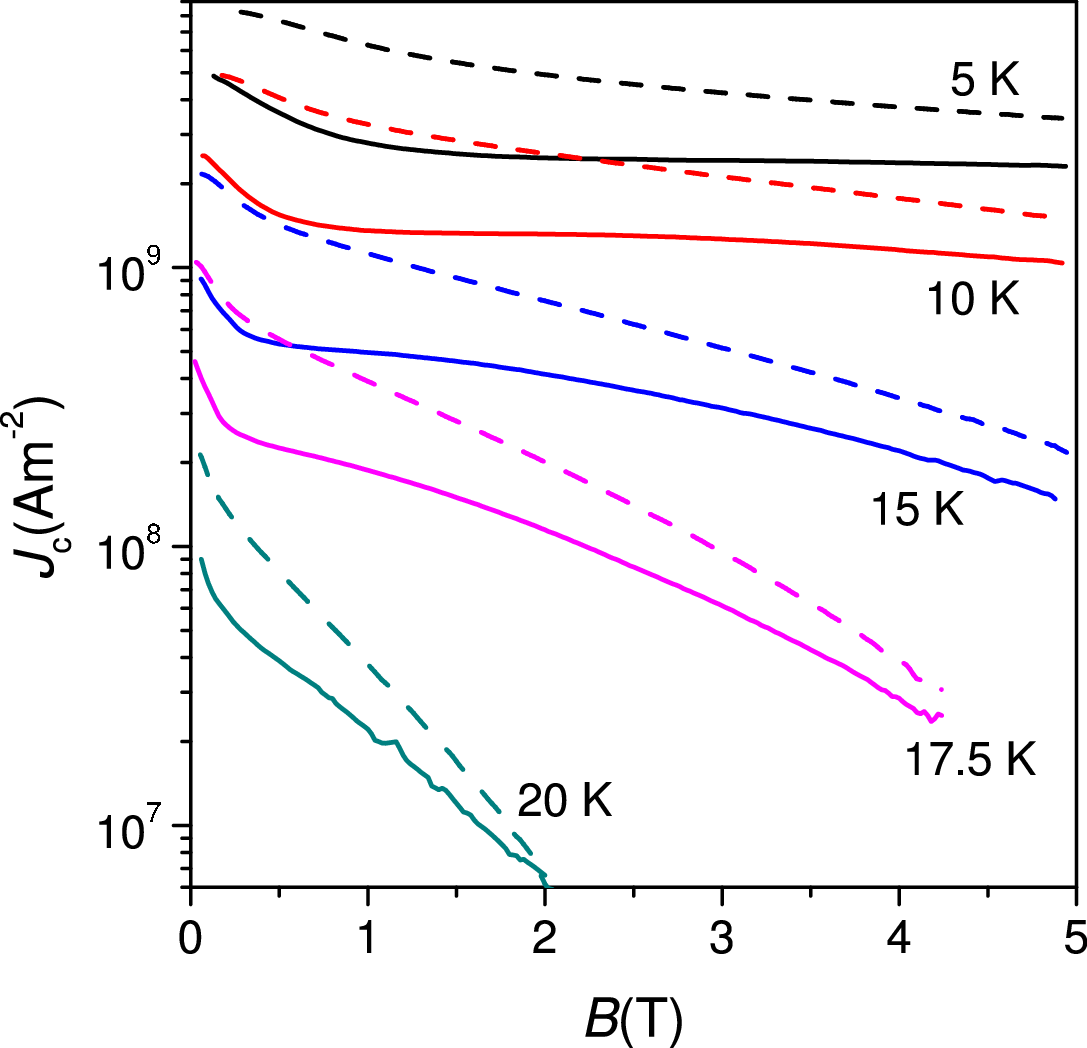}
    \caption{\label{fig:jc} (Color online) Critical current density within the $ab$ plane before
(solid lines) and after (dashed lines) neutron irradiation as a
function of magnetic induction.}
\end{figure}

The critical current densities, $J_\mathrm{c}(B)$, obtained from
magnetization measurements at temperatures from 5 to 20\,K are
presented in Fig.~\ref{fig:jc} (solid lines) and
confirm\cite{Yam09} that the major part of the superconducting
phase diagram is dominated by irreversible properties, in contrast
to many HTSC and LTSC single crystals. This indicates that Ba-122
crystals contain a rather effective as-grown pinning matrix. Very
recently, STM investigations of similar crystals\cite{Yin08}
revealed impurities with a radius of about 1 - 2\,nm (according to
the figures of Ref.~\onlinecite{Yin08}), which are presumably
related to Co or Fe vacancies. Their size would perfectly match to
the coherence length of the material ($\sim 2 - 3$\,nm) and could
therefore provide perfect core pinning. Controlling their density,
e.g. during the growth process, could make the material very
attractive for applications. After a sharp decay of $J_\mathrm{c}$
at low fields, the curves are almost constant over a larger field
interval and show traces of a fishtail effect (but no real second
maximum - cf. Fig.~\ref{fig:jc} solid lines). According to theory,
the sharp decay at low fields reflects the significant role of the
repulsive vortex vortex interaction (i.e. the elastic energy) at
that region, whereas the fishtail \cite{Mik01} results from a more
pronounced role of the pinning energy at higher fields, which
competes with the elastic (vortex) energy and leads to a more
disordered lattice and a better adjustment to the defect matrix.

\begin{figure}
    \centering
    \includegraphics[clip, width =0.35\textwidth]{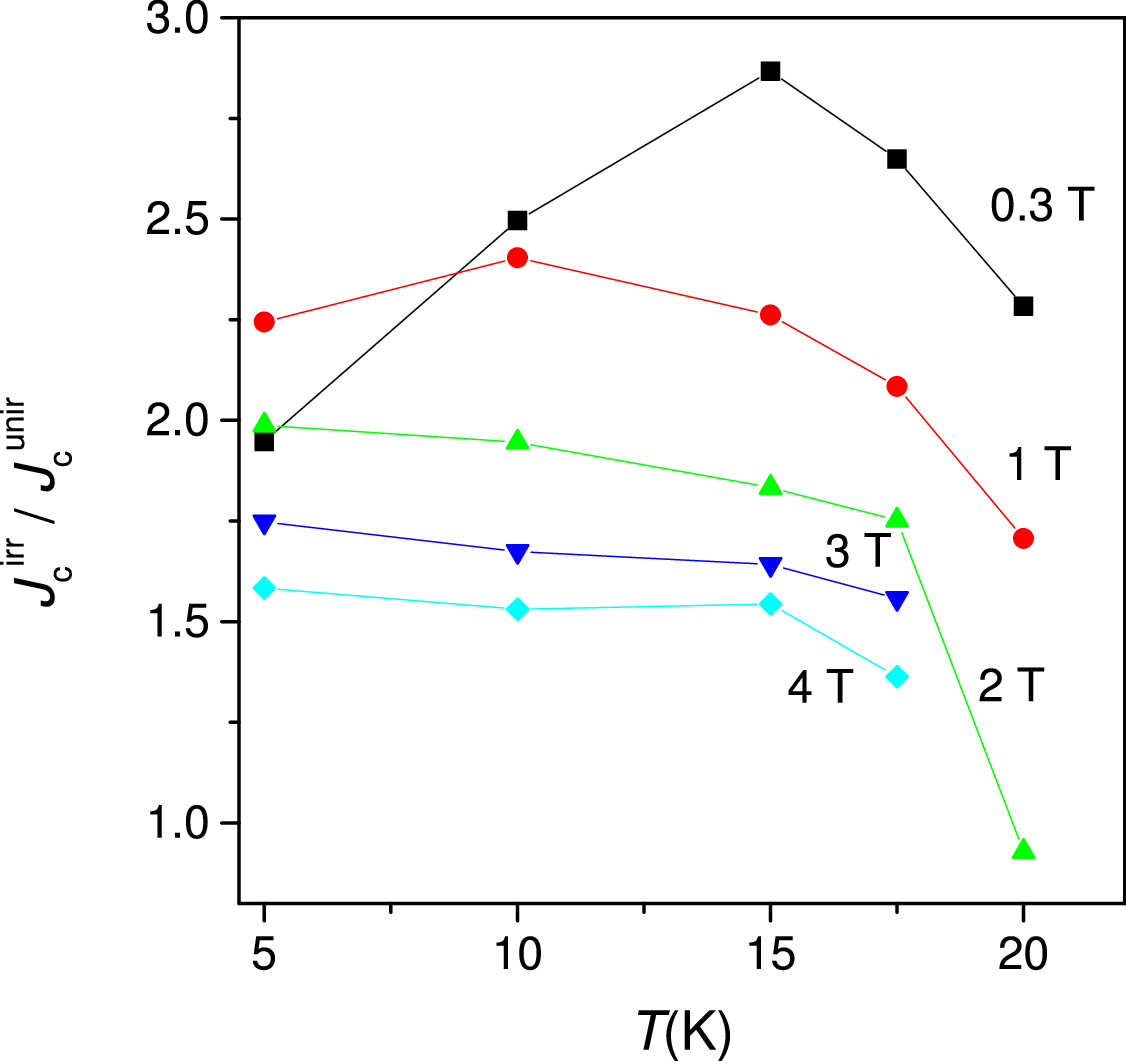}
    \caption{\label{fig:Enhancement} (Color online) Ratio of critical currents before
($J_\mathrm{c}^\mathrm{unir}$) and after
($J_\mathrm{c}^\mathrm{irr}$) neutron irradiation as a function of
temperature at various applied fields.}
\end{figure}

Neutron irradiation enhances $J_\mathrm{c}$ - Fig.~\ref{fig:jc}
(dashed lines). Since the reversible properties are only slightly
affected by the chosen neutron fluence, the modification of the
pinning matrix seems to be mainly responsible for these effects.
Neutron irradiation is known to introduce very effective pinning
centers in a wide range of superconducting materials including the
cuprates like Y-123 \cite{Sau98} and MgB$_2$ \cite{Zeh04}, where
spherical defects with a size comparable to the coherence length
are introduced. The pronounced $J_\mathrm{c}$ enhancement in
Ba-122 suggests a similar defect generation, but note that the
enhancement is smaller than in many other single crystals. This is
in agreement with our assumption of a rather effective as-grown
defect matrix (compared to other materials), since the relative
effect of strong (radiation induced) pinning centers is expected
to be larger in very clean samples \cite{Zeh02b}. For instance,
Y-123 exhibits a strongly temperature dependent enhancement
\cite{Zeh02b} by up to a factor of 100 at high temperatures (e.g.
at 77\,K) and roughly 4 - 5 at low temperatures (5\,K, all data
refer to low fields) which can be explained by the thermal
fluctuations and the more two dimensional character at high
temperatures. The enhancement in MgB$_2$ single crystals
\cite{Zeh04} was found to be about 5 - 7 at low fields and to
remain quite constant over the temperature range.

The results on Ba-122 are presented in Fig.~\ref{fig:Enhancement}.
The enhancement of the critical current density $ J_\mathrm{c,irr}
/J_\mathrm{c,unir}$ lies between about 3 and 1. Like in MgB$_2$
there is only a weak temperature dependence, which suggests that
both the as-grown and the radiation induced pinning sites are not
much smaller than the coherence length, since very small pinning
sites would be effective only at low temperatures, where the
coherence length is small and the thermal energy negligible. The
traces of the fishtail effect observed in the as-grown state
disappear upon irradiation indicating that the new pinning sites
transfer the ordered vortex lattice at low fields into a pinning
dominated disordered phase.  This behavior agrees with theory and
was found in many HTSC samples upon irradiation
\cite{Zeh02b,Wer00} and recently also in Sm-1111 \cite{Eis09b}.
Accordingly, the enhancement is largest at those fields, where the
ordered state was mostly pronounced in the unirradiated state. At
higher fields, the enhancement decreases and approaches 1 near the
irreversibility field in the high temperature measurements (where
sufficiently high fields are accessible). Similar effects have
been found in many Y-123 and other HTSC crystals, where the
irreversibility line is not significantly affected by neutron
irradiation.

\section{Conclusions}

In conclusion, we found that electron scattering increases after
neutron irradiation leading to a slight decrease of $T_\mathrm{c}$
and the anisotropy. The critical current density increases by a
factor of up to 3, indicating that the radiation induced defects
are effective pinning centres. These results are in qualitative
agreement with our recent study on polycrystalline Sm-1111
\cite{Eis09b}.

\begin{acknowledgments}
We are grateful to D. C. Larbalestier, A. Gurevich, M. Putti, D.
V. Abraimov, C. Tarantini and F. Kametanni for discussions. Work
at the NHMFL was supported by AFOSR, NSF, and the State of
Florida.
\end{acknowledgments}


\end{document}